# On the effect of Re addition on microstructural evolution of a CoNi-based superalloy

P. Pandey[1*], A. K. Sawant[1], B. Nithin[1], Z. Peng[2], S. K. Makineni[1,2*], B. Gault[2], K. Chattopadhyay[1*]

[1]Department of Materials Engineering, Indian Institute of Science, Bangalore 560012, India.

[2]Department of Microstructure Physics and Alloy Design, Max-Planck-Institute für Eisenforschung, 40237 Düsseldorf, Germany

*Corresponding Authors: prafull1011@gmail.com (PP), sk.makineni@mpie.de (SKM), kamanio@iisc.ac.in (KC)

**Abstract**

In this study, the effect of rhenium (Re) addition on microstructural evolution of a new low-density Co-Ni-Al-Mo-Nb based superalloy is presented. Addition of Re significantly influences the γ' precipitate morphology, the γ/γ' lattice misfit and the γ/γ' microstructural stability during long term aging. An addition of 2 at.% Re to a Co-30Ni-10Al-5Mo-2Nb (all in at.%) alloy, aged at 900°C for 50 hours, reduces the γ/γ' lattice misfit by ~ 40% (from +0.32% to +0.19%, measured at room temperature) and hence alters the γ' morphology from cuboidal to round-cornered cuboidal precipitates. The composition profiles across the γ/γ' interface by atom probe tomography (APT) reveals Re partitions to the γ phase ($K_{Re} = 0.34$) and also results in the partitioning reversal of Mo to the γ phase ($K_{Mo} = 0.90$) from the γ' precipitate. An inhomogeneous distribution of Gibbsian interfacial excess for the solute Re ($\Gamma_{Re}$, ranging from 0.8 to 9.6 atom.nm$^{-2}$) has been observed at the γ/γ' interface. A coarsening study at 900°C (up to 1000 hours) suggests that the coarsening of γ' precipitates occurs solely by an evaporation–condensation (EC) mechanism. This is contrary to that observed in the Co-30Ni-10Al-5Mo-2Nb alloy as well as in some of the Ni-Al based and high mass density Co-Al-W based superalloys, where γ' precipitates coarsen by coagulation/coalescence mechanism with extensive alignment of γ' along <100> directions as a sign of microstructural instability. The γ' coarsening rate exponent ($K_r$) and γ/γ' interfacial energy are estimated to be $1.41 \times 10^{-27}$ m$^3$/s and 8.4 mJ/m$^2$, which are comparable and lower than Co-Al-W based superalloys.

***Keywords***: Rhenium (Re); Cobalt-based superalloys; Lattice misfit; Atom probe tomography (APT); Coarsening kinetics.

## 1. Introduction

γ/γ' based superalloys are a class of high temperature alloys that are critical for fabricating parts and components in the aero- and land-based gas turbine engines [1–3]. In these alloys, the high temperature strength and the microstructural stability are achieved through γ' precipitates having $L1_2$ ordered structure coherently embedded in the disordered face-centered-cubic (fcc) γ matrix [2,4]. This characteristic γ/γ' microstructure is seen in Ni-Al and Co-Al based systems [5–8]. Commercial Ni-Al based superalloys contain 8 to 12 alloying elements that provide an excellent combination of physical, mechanical and corrosion/oxidation properties even at temperatures equal or greater than 1100°C [2,9–13]. The addition of elements with high degree of γ matrix solid solubility such as rhenium (Re), ruthenium (Ru), tungsten (W) and molybdenum (Mo) have led to the development of multiple generations of these superalloys [2,14–18]. However, if added beyond their solubility limit, detrimental topologically close-packed (TCP) phases form [3,19,20]. Re was shown to be most effective in enhancing high temperature mechanical properties of Ni-Al based superalloys [21–23]. In these alloys, Re partitions to the γ matrix [24] which is supported by first-principles calculations [25]. Additionally, Re is also known to segregate at the γ/γ' interfaces and slows down the coarsening kinetics of γ' precipitates by reducing the lattice mismatch [26–29] and onset of creep at elevated temperatures [15,30].

The recently developed Co-Al based alloys with the γ/γ' microstructure have demonstrated comparable high temperature creep than the second generation Ni-Al based superalloys [31–33]. This class can be divided into Co-Al-W based [5,6] and low density Co-Al-Mo-Nb/Ta based superalloys [8,34–36]. The microstructure of the former contains $Co_3(Al,W)$ γ' precipitates [37,38] while the latter has $Co_3(Al, Mo, Nb/Ta)$ γ' precipitates [34,35] coherently embedded in the γ-Co matrix. These precipitates are metastable, only exist in a narrow phase field region and were shown to decompose into the equilibrium phases at high temperatures [39–41]. The research efforts on these alloys, covering aspects of multiple element additions and their effect on mechanical [42–44] and oxidation/corrosion properties [45–48] have recently gained momentum. In particular, Ni addition was shown to stabilize γ', increase the γ/γ' phase-field, solvus temperature and enhance the solubility limit of refractory elements/alloying additions in the alloy [49–53].

The effect of Re addition to these alloys is limited [54,55]. Recent experimental and first principles kinetic Monte Carlo studies by Neumeier et al. show Re is the slowest diffusing element amongst all transition metals in fcc-Co [56]. Due to the higher vacancy formation energy in Co, the diffusivity and activation energy barrier of Re is greater than Ni [56,57]. However, Kolb et al. [54] and Volz et al. [55] showed Re addition to a single crystalline Co-Ni-Al-W based alloy leads to a marginal or no improvement in creep properties compared to the role of Re in the $2^{nd}$ and $3^{rd}$ generation Ni-Al based superalloys. This was attributed to the low partitioning of Re to the γ matrix phase. Hence, it was concluded that Re is not crucial for Co-Al-W based superalloys.

The addition of Re could have an important role for the long term microstructural stability of Co-Al based superalloys. However, the role of Re in the coarsening behavior of cobalt based alloys, in particular that of the low-density Co-Al-Mo-Nb/Ta based superalloys, has not been explored rigorously. Some studies on the kinetics and associated models have been reported for Re added Ni-Al based superalloys [27,28,58]. They show that the coarsening of precipitates follows Lifshitz-Slyozov-Wagner (LSW) based Ostwald ripening mechanism with slight variation in temporal exponents.

The present work reports a detailed study on the effect of Re addition on a low density W-free γ/γ′ Co-Al based superalloys. Microstructural evolution, mechanical properties at high temperatures and γ' coarsening kinetics have been elucidated. Advanced characterization tools such as atom probe tomography (APT) for atomic compositional partitioning across γ/γ' and scanning/transmission electron microscopy (SEM and TEM) have been used for measuring the temporal evolution of γ' precipitate size and number density. Subsequently, modified LSW models for concentrated multicomponent alloy have been validated in terms of temporal exponents and coarsening rate constants. These were compared with reported results for Ni-Al based and Co-Al-W based superalloys.

## 2. Experimental
### 2.1 Alloy preparation

The alloy ingot of 50 grams with the nominal composition Co-30Ni-10Al-5Mo-2Nb-2Re (all in at.% and named as 2Nb2Re) was prepared using pure metals (99.99% purity) in a vacuum arc-melting unit equipped with water cooled copper hearth and a tungsten electrode under argon atmosphere. The alloy was melted 8 times with flipping after each cycle for homogenous melting throughout the ingot. The melted alloy ingot was cast in the form of rods (diameter, ϕ-3 mm, length, L-80 mm) under argon atmosphere using a split water-cooled copper mold in a vacuum suction casting unit. The cast rods were solution-treated at 1250 °C for 20 h in a tube furnace operating under a vacuum of $10^{-3}$ Pa followed by water quenching. The solutionized cast samples were aged at 900 °C for 50 h, 200 h, 300 h, 500 h and 1000 h followed by water quenching for studies on temporal evolution of the γ' precipitates. The composition of the alloy was measured by energy dispersive spectroscopy (EDS) and it is shown in Table 1.

*Table 1:* *Nominal and measured composition (in atomic %) of studied alloy as measured by EDS in SEM*

| Nomenclature | Nominal Composition (at.%) | Measured Composition (at.%) |
|---|---|---|
| 2Nb2Re | Co-30Ni-10Al-5Mo-2Nb-2Re | Co-30.4±0.1Ni-9.5±0.3Al-5.3±0.1Mo-2.4±0.2Nb-2.0±0.1Re |

**2.2 Scanning and transmission electron microscopy (SEM and TEM)**

The microstructures of the heat-treated alloy samples were observed in a scanning electron microscope (SEM, Helios NanoLab 460) equipped with a field-emission gun (FEG) source. The samples were polished with standard metallography procedure using 800 to 3000 Si grit papers. The final sample surface preparation was carried out by chemo-mechanical polishing technique using colloidal silica in vibratory polisher (VibroMet™2, Buehler). Transmission electron microscopy (TEM) for the heat-treated samples was carried out using an FEG TEM microscope (Tecnai F-30, FEI) operated at 300 kV. For TEM samples, disk-shaped specimens were prepared from the heat treated cast 3 mm rods cut in 1mm-thick slices and subsequently mechanically polished to a thickness less than 100 μm. These discs were electro-polished using an electrolyte of 5% perchloric acid in methanol solution at a temperature of -38 °C and a voltage of 16 V to get the transparent regions for TEM studies.

**2.3 X-Ray diffraction (XRD)**

The X-ray diffraction measurements for the determination of the lattice misfit between γ and γ' were performed at room temperature using a Rigaku Smart lab high-resolution X-ray diffractometer with Cu Kα radiation equipped with Johansson optics that eliminates the contribution from Kα$_2$ and Kβ components from the source. The asymmetric 200 reflections from the γ and γ' phases have been obtained by performing asymmetric θ−2θ scans using a four-circle goniometer in a Eulerian cradle. Prior to the scan, the alignment conditions for the in-plane rotation and lattice inclination (out-of-plane) were determined by performing phi ($\Phi$) and chi ($\chi$) scans, respectively. The data was collected for 0°-360° with a step of 3° phi ($\Phi$) and tilting 52.5°-57.5° with a step of 0.1° in chi ($\chi$) range. The $\theta$–$2\theta$ scan was collected at an angular range of 50°- 53° with a scan step of 0.001° and time per step of 1°/minute. The peak positions corresponding to the $\gamma$ and $\gamma'$ phases were determined by deconvolution of the measured diffraction profile using a pseudo-Voigt function in a Fityk peak analysis program. Among all the possible combination of peak fits, one with best $R^2$ value was chosen.

## 2.4 Mechanical and thermophysical properties

The compression tests for the aged (900°C, 50 hours) samples were carried out in a servo-hydraulic driven universal testing system (UTS, DARTEC make) under a constant strain rate of $10^{-3}$ $s^{-1}$ and at temperatures of 25 °C, 470 °C, 570 °C, 670 °C, 770 °C, 870 °C and 970 °C. The cylindrical samples for the compression tests were cut from the 3mm diameter rods by keeping the height/diameter (H/D) ratio 1.5 to 2.0.

The transformation temperatures for the heat treated 2Nb2Re alloy were determined using differential scanning calorimetry (DSC, STA449F3 NETSCH make). Samples weighing 80 to 100 mg were cut from the heat treated rods and heated at a rate of 10 °C/min till 1450 °C followed by cooling to room temperature at the same rate. The $\gamma'$ dissolution temperature was determined from the point of intersection between extrapolated base line and inflectional tangent at the beginning of the dissolution peak in the DSC heating curve.

The density of the alloy was measured at room temperature following the ASTM standard (ASTM 311-17) which is based on Archimedes principles. Prior to the measurement, the sample weighing ~ 10 grams was soaked overnight in the acetone and subsequently dried.

## 2.5 Atom Probe Tomography

The atomic scale compositional measurements for the heat-treated 2Nb2Re samples were carried out using atom probe tomography (APT). Specimens for the APT measurements were extracted from the polished samples using a dual-beam SEM/focused-ion-beam instrument (FIB, Helios Nanolab 600) in conjunction with an in-situ lift-out protocol described in the reference [59,60].

The Ga ion milling was carried out in multiple steps at 30 kV with reduced ion current in each successive step to obtain a sharp needle. Finally, the cleaning of the prepared tips at 2 kV with 8 pA current was carried out to remove the severely damaged regions due to Ga ion source. APT measurements were performed in LEAP™ 5000XR (Cameca instruments Inc., Madison, WI) in laser-pulsing mode. The ultraviolet picosecond laser pulses were applied at a repetition rate of 125 kHz with a pulse energy of 50 pJ. The detection rate was maintained at 5 detection events per 100 pulses. The specimen's base temperature was kept at 50 K. The data analysis was performed using the IVAS™ 3.8.0 software package. In addition, an in-house APT data treatment routine developed in Mathworks MATLAB was applied to accurately quantify the segregation of solutes at the γ/γ' interface. More details about this routine can be found in the reference [61].

**2.6 Quantitative stereological measurements**

The γ' precipitate sizes ($<r>$) and their volume fraction ($\phi^{\gamma'}(t)$) were measured from SEM images of [100] oriented grains using the imageJ (https://imagej.nih.gov/ij/) software. At least 500 precipitates from different areas of each sample were taken for measurements. The precipitate size calculation is based on an area equivalent circle radius and in which the calculated precipitate area is considered equivalent to the circle area by the equation

$$A_p = \pi * r_p^2 \quad (1)$$

Where, $A_p$ is the calculated precipitate area and $r_p$ is the equivalent circle radius of the precipitate

The average precipitate size <r> was calculated from the precipitate size distribution which fits well with the normal size distribution. The error in the average precipitate size measurement is calculated by the relation as follows

$$\sigma_r = \frac{\sigma_{std}}{\sqrt{N}} \quad (2)$$

Where, $\sigma_r$ is the error in the precipitate size ($<r>$) measurement, $\sigma_{std}$ is the standard deviation of the average precipitate size (width of the precipitate size distribution) and N is the no. of precipitates considered for the average precipitate size measurement. Equation (2) indicates that the error in measurement decreases on increasing the number of precipitates considered for measurement.

The precipitate volume fraction [$\phi'(t)$] was calculated stereologically for each aging condition by using line intercept method in at least 5 SEM micrographs which was acquired from 5 different [100] oriented grains. The calculations of 5 micrographs is finally averaged out and reported along with standard deviation. The areal number density [$N_A(t)$] was calculated by counting the number of precipitates and dividing it by the total area using image J software. The precipitate counting was based on following method: (a) A precipitate confined within the micrograph is counted as one (b) the partially confined precipitate is counted as 0.5 and (c) The precipitates that are joined through a neck are counted as 2. The precipitate number density [$(N_v(t)$] is calculated using quantitate stereology measurement based relation which can be given as [62]

$$N_v = \frac{N_A}{2<r>} \qquad (3)$$

Where, $N_v$ is the precipitate number density and $<r>$ is the average precipitate radius

The average precipitate center to center distance [$<\lambda_{c-c}(t)>$] is determined by the relationship

$$<\lambda_{c-c}> = 2 * \left(\frac{3}{2\pi N_v}\right)^{\frac{1}{3}} \qquad (4)$$

The average precipitate edge to edge distance [$<\lambda_{e-e}(t)>$] can be calculated as follows

$$<\lambda_{e-e}> = 2 * \left[\left(\frac{3}{2\pi N_v}\right)^{\frac{1}{3}} - <r>\right] \qquad (5)$$

**2.7 Coarsening model**

We first briefly introduce the coarsening model adopted herein. The classical matrix diffusion limited precipitate coarsening model proposed by separate works of Lifshitz, Sylozov and Wagner (commonly known as LSW model) [63,64]. It is based on the Ostwald ripening due to the capillary effect that is related to the Gibbs-Thompson equation. The driving force for the coarsening develops due to the process of minimization of the total interfacial energy between the precipitate and the matrix. This minimization leads to the growth of larger particles at the expense of smaller particles through an evaporation-condensation (EC) mechanism [65]. However, the LSW theory for quasi-steady-state assumes the following: the matrix is dilute, the precipitate volume fraction is negligible, there is no elastic interaction between the precipitates, and the spherical shaped precipitates are randomly distributed in the matrix with no diffusional field overlap. The coarsening occurs through volume diffusion-controlled EC, which does not take into account the coagulation or coalescence of the precipitates. Further it is assumed that the compositions of the precipitates and the matrix are in a quasi-steady state. Based on these assumptions, the solution of the continuity equation, at t → ∞ predicts three asymptotic solutions of the temporal power law exponent:

1) $<r> \propto t^{1/3}$, where $<r>$ is average precipitate size and $t$ is time. (6)
2) $N_v(t) \propto t^{-1}$, where $N_v(t)$ is precipitate number density. (7)
3) $\Delta C_\gamma(t) \propto t^{-1/3}$, where $\Delta C_\gamma(t)$ is matrix supersaturation. (8)

Typically, superalloys are multicomponent alloys with non-dilute phases and have precipitate volume fraction often approaching 70 to 80%. Several modifications as well as extended models to the LSW theory have been proposed to take care of the assumptions on the precipitate volume fraction and multicomponent non-dilute solutions [66–71]. All these models predict the asymptotic solution in the steady-state region at t → ∞, similar to the LSW solutions (eq. 6, 7, and 8). However, with the modification in the amplitude of the temporal exponent (*n*) and the coarsening kinetic rate, a more generalized solutions on the concentrated, non-ideal system for the evolution of the γ/γ' microstucture can be written as follows [69]:

1) $<r(t)>^n - <r(0)>^n = K_r t$, where $K_r = \dfrac{8DV_m^{\gamma'} C_{eq}^{\gamma}(1-C_{eq}^{\gamma})\sigma}{9RT(C_{eq}^{\gamma'}-C_{eq}^{\gamma})^2}$ (9)

$<r(t)>$ and $<r(0)>$ are the average precipitate size at time $t$ and at the onset of coarsening. $K_r$ is the precipitate coarsening rate constant, $\sigma$ is the interfacial energy between γ and γ', D is the diffusivity of the slowest diffusing species in the matrix, $V_m^{\gamma'}$ is the molar volume of γ′ precipitate, $C_{eq}^{\gamma}$ is the equilibrium solubility of solute in the γ matrix, $C_{eq}^{\gamma'}$ is the equilibrium solubility of solute in the γ′ precipitate, R is the universal gas constant, and T is the absolute temperature.

2) $N_v(t) \cong 0.21 \dfrac{\phi^{eq}}{K_r} t^n$, (10)

where, $\phi^{eq}$ is the equilibrium volume fraction of γ′ precipitates

3) $\Delta C_i^\gamma = [(<C_i^\gamma(t)>) - (<C_i^\gamma(\infty)>)] = K_i t^n$, (11)

where, $\Delta C_i^\gamma$ is the difference between the concentration of an element $i$ in matrix at time $t$, $(<C_i^\gamma(t)>)$ and the average equilibrium concentration$(<C_i^\gamma(\infty)>)$. $K_i$ is the coarsening rate constant for the matrix supersaturation. In equation (9), (10), and (11), an arbitrary temporal exponent $n$ appears. In the present work, the coarsening kinetic studies were carried out based on the first two asymptotic solutions (equation 9 and 10). We would not be able to take into account the third solution (equation 11) since our measurements on compositions of γ matrix phase and γ' precipitate did not change significantly even during long term aging at 900°C up to 1000 hours.

## 3. Results
## 3.1 Microstructural, mechanical and thermophysical properties of the alloy after solutionising and aging at 900°C, 50 hours

### 3.1.1 *Microstructure of the alloy after solutionising at 1250°C*

Fig. 1(a) shows an overview image of the alloy after solutionising at 1250°C for 20 hours. The microstructure contains equiaxed grains with 50 to 300-micron size range and contains annealing micro-twins within some of the grains.

### 3.1.2 *Microstructure of the alloy after aging at 900°C, 50 hours*

Fig. 1(b) shows a secondary electron micrograph for the alloy aged at 900°C for 50 hours. The image was taken near to <100> axis. It shows a homogenous distribution of cuboidal precipitates with rounder corners, as readily visible in the highly magnified micrograph in Fig. 1(c). Figure 1(d) shows a TEM darkfield (DF) image taken from a (010) superlattice reflection along the [100] cubic zone axis. The diffraction pattern (DP) is shown as an inset in the same figure. The DP reveals presence of superlattice spots corresponding to the γ' $L1_2$ ordering along with the main γ fcc matrix reflections. The DF highlights uniform cuboidal γ' precipitates with the rounded corners embedded in the dark γ matrix. This γ' morphology is distinct from the alloy without Re i.e. Co-30Ni-10Al-5Mo-2Nb alloy (named as 2Nb in the subsequent sections) that consists of cuboidal γ' precipitates with sharp corners [34,72]. The γ′ volume fraction ($\phi_t^{\gamma'}$) in 2Nb2Re is measured to be ~ 53 ± 3% and is similar to the value obtained in 2Nb alloy [72].

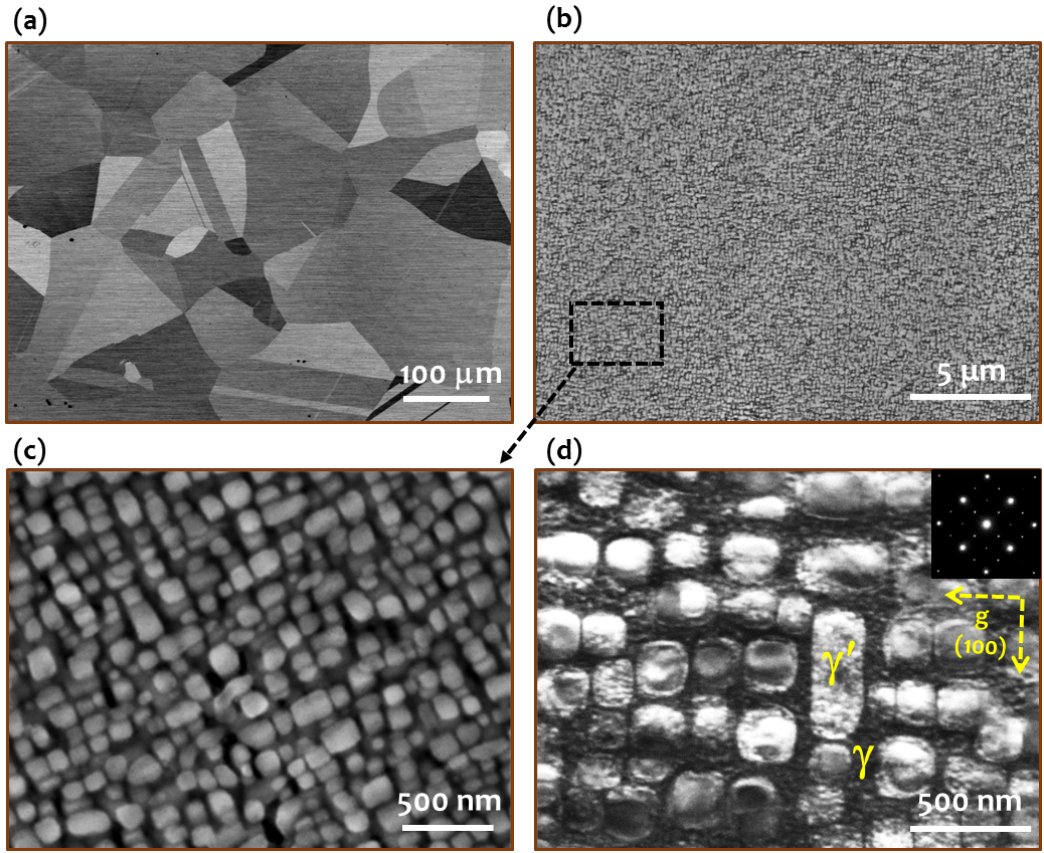

*Figure 1:* *(a) Back scattered and secondary electron images for Co-30Ni-10Al-5Mo- 2Nb-2Re (2Nb2Re) alloy (a) after solutionising heat treatment at 1250 °C for 20 h followed by water quenching (b-c) aged at 900 °C for 50 hours, showing uniformly distributed γ′ precipitates in the γ matrix. (d) Transmission electron microscopy (TEM) centered darkfield image taken near to [100] cubic zone axis acquired from 010 superlattice reflection showing distribution of coherent γ′ precipitates in the γ matrix. The [100] diffraction pattern is also shown as an inset.*

3.1.3  *Compositional partitioning across the γ/γ′ interface for the aged alloy*

Fig. 2(a) shows the APT reconstruction of a specimen extracted from the 2Nb2Re alloy subjected to aging at 900 °C for 50 hours. The γ and γ′ phase can be identified by the solute partitioning. The γ/γ′ interfaces are delineated by iso-composition surfaces with a threshold of 46 at.%Co. The Nb atoms, shown in green, partition to γ′ precipitates while the Re atoms, shown in red, partitions to γ. Figure 2(b) shows the composition profile across the γ/γ′ interface for the solutes Al, Mo, Nb and Re. The Co and Ni compositional profile are shown in supplementary Fig. S1. The quantitative partitioning values across the γ/γ′ for each solute were calculated by the formula

$$K_i = C_i^{\gamma'}/C_i^{\gamma} \qquad (12)$$

where, $K_i$ is the partitioning coefficient of element $i$, $C_i^{\gamma'}$ and $C_i^{\gamma}$ is the composition of element $i$ in $\gamma'$ and $\gamma$. The compositions and $K_i$ values are listed in Table 2. Nb, Al and Ni strongly partition to $\gamma'$ with $K_{Ni}$ = 1.39, $K_{Nb}$ = 6.57 and $K_{Al}$ = 2.82, respectively. Similar partitioning behavior of these solutes is shown in the 2Nb alloy [34]. However, we observe a reverse partitioning of Mo i.e. it partitions to $\gamma$ phase with respect to $\gamma'$ precipitate in the present 2Nb2Re Alloy. The measured partitioning coefficient for Mo is $K_{Mo}$= 0.90. The Re strongly partitions to $\gamma$ with a partitioning value $K_{Re}$ = 0.34. A careful observation at the $\gamma/\gamma'$ interface shows that the solutes Re and Mo segregates at the $\gamma/\gamma'$ interface as depicted in Fig. 2(c).

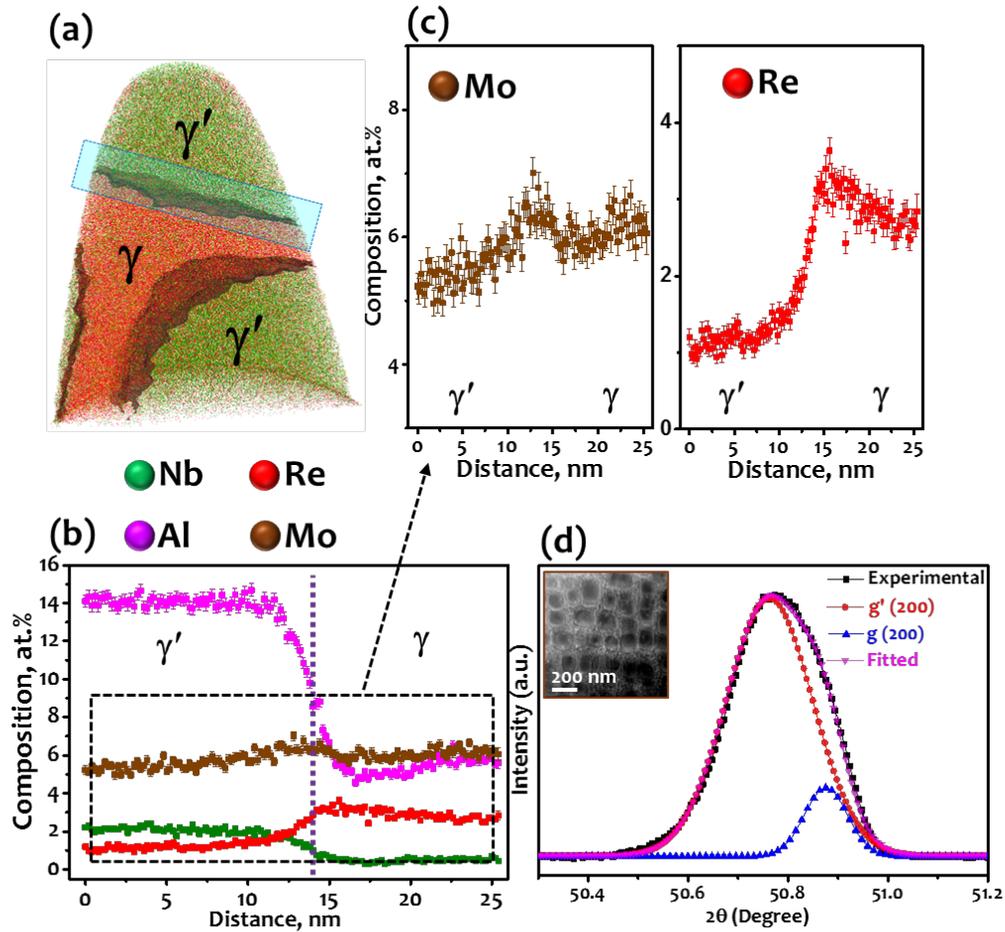

*Figure 2: (a) 3D APT reconstruction of the tip from the 2Nb2Re sample aged at 900 °C for 50 hours. The γ/γ' interfaces are shown as iso-composition surfaces at 46 at.% Co. (b) The compositional profile across the γ/γ' interface for the solutes Al, Nb, Mo and Re. (c) The magnified view of the Mo and Re compositional profile across the interface showing evidence of significant*

*interfacial segregation. (d) X-ray diffraction peak (asymmetric (200) plane reflection) for the 900 °C for 50 hours aged 2Nb2Re alloy showing two overlapping peaks corresponding to γ matrix and γ′ precipitate. The inset shows a STEM-HAADF image taken near to [100] cubic direction.*

3.1.4 *Evaluation of Lattice mismatch across γ/γ′ interface*

High resolution X-ray diffraction was used to measure the γ/γ′ lattice misfit at room temperature by capturing (200) reflection for the aged 2Nb2Re alloy. The asymmetric diffraction peak is plotted and shown in Fig. 2(d). The asymmetric nature of the peak is due to the lattice misfit at the γ/γ′ interface and the tetragonal distortion of the γ matrix caused by the presence of coherency stresses at the γ/γ′ interfaces [73]. The deconvolution (pseudo-voigt function fit) of the diffraction peak shows two overleaping peaks that corresponds to the γ matrix and the γ′ precipitate. The peak intensities cannot be compared quantitatively with the volume fraction of phases due to possible influences of other factors such as ordering parameters and texture component. However, due to known positive lattice misfit in the Co-based superalloys [6,72,74], it is safe to assume that the larger intensity peak at lower 2θ value corresponds to γ′ precipitates with a higher volume fraction and smaller intensity peak at higher 2θ value corresponds to the γ matrix. The constrained bulk lattice misfit between γ′ and γ was calculated by the formula

$$\delta = \frac{2(a_{\gamma\prime} - a_\gamma)}{(a_{\gamma\prime} + a_\gamma)} \quad (13)$$

where, $a_{\gamma\prime}$ is the lattice parameter of the γ′ precipitate and $a_\gamma$ is the lattice parameter of the γ phase. As the lattice parameter of γ′ (3.594 Å) is more than the γ matrix (3.587 Å), the calculated constrained lattice misfit for the 2Nb2Re alloy is positive with the value of $\delta$ = +0.19%. The Re addition to 2Nb alloy ($\delta$ = +0.32%) lowers the lattice misfit by 40% [72]. This reduction in the lattice misfit after the addition of Re is reflected in the change in γ′ morphology from cuboidal with sharp corners to round-cornered cuboidal shape of γ′ precipitates as shown in Figures 1(b-d) and in the inset of Figure 2(d) that shows an Scanning-TEM high-annular-angle-dark-field (HAADF) image along [100] cubic zone.

### 3.1.5 *Thermophysical properties of the aged alloy*

Fig. 3(a) shows the DSC heating curve for the 2Nb2Re alloy aged at 900°C for 50 hours. The peaks that appeared in the heating curve have been marked as P1 and P2. The onset of the first peak, P1, during heating corresponds to the γ' dissolution temperature (solvus). P2 peak is solidus temperature of the 2Nb2Re alloy. The γ' solvus temperature was measured to be ~ 1000°C which is higher by 10°C from the value reported for 2Nb alloy (990°C). Additionally, the solidus temperature of the 2Nb2Re alloy is also higher by 10°C. However, in Co-Al-W-based alloys, Re is shown to have no effect on the γ' solvus temperature [42].

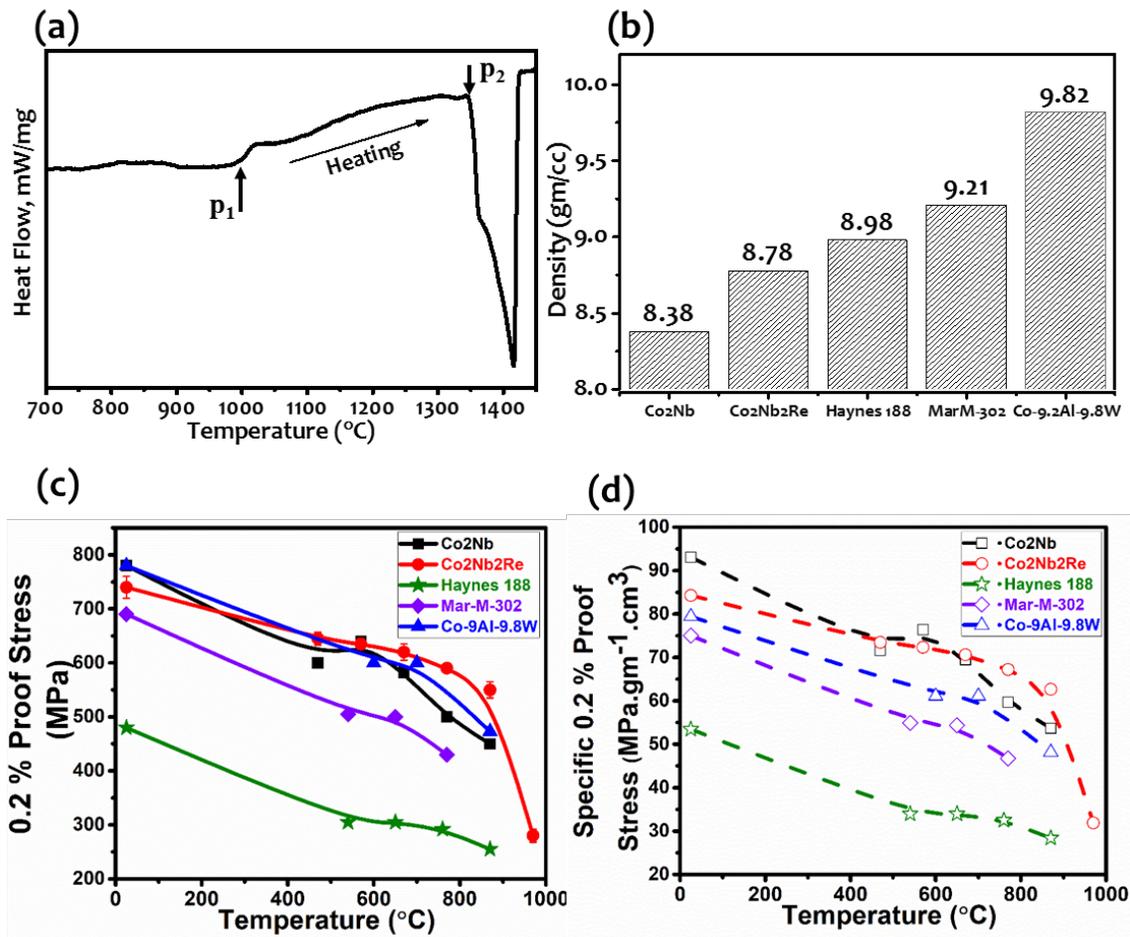

*Figure 3: (a) The DSC heating curve for the 2Nb2Re alloy aged at 900°C for 50 hours showing onset of γ' dissolution temperature (γ' solvus temperature) and incipient melting of alloy. (b) The density comparison of present 2Nb2Re with base 2Nb alloy and other commercially available Co based superalloys. (c) 0.2% proof stress vs. temperature and (d) 0.2% specific proof stress vs.*

*temperature plot for aged 2Nb2Re alloy and its comparison with other commercially available Co based superalloys.*

The measured density of the 2Nb2Re alloy is 8.78 gm/cm$^3$, higher than the 2Nb alloy (8.38 gm/cc). Figure 3(b) shows the comparison plot with other Co based superalloys. Even though the density is increased by Re addition to the base 2Nb alloy, it is still lighter than several other Co based superalloys.

### 3.1.6 *Mechanical properties of the aged alloy*

The 0.2% proof stress (PS) values for the 2Nb2Re alloy were evaluated by compression testing at temperatures in the range between 25 °C and 970 °C. Fig. 3(c) shows plots of 0.2% PS (proof stress) vs temperature for the 2Nb2Re, 2Nb and other Co-based superalloys. We observe a monotonous drop in 0.2% PS from a value of 750 MPa at room temperature to 550 MPa at 870 °C and further reduces to 300 MPa at 970 °C. The Specific 0.2% PS values were also calculated and compared in Fig. 3(d). The 2Nb2Re alloy shows similar room temperature and high temperature values to that reported for the base alloy 2Nb and the Co-Al-W alloys. However, they are superior to several other commercial Co based alloys. Due to a relatively lower density, the 2Nb2Re alloy exhibits specific 0.2% PS values significantly higher than Co-Al-W alloys as well as other Co-based solid solution alloys at all the temperatures. The sudden drop in 0.2%PS at 970 °C for the 2Nb2Re alloy can be attributed to the dissolution of the strengthening γ' precipitates since the temperature is close to the γ' solvus temperature (1000 °C).

## 3.2 Microstructural evolution of the 2Nb2Re alloy during long term aging at 900°C

### 3.2.1 *γ' precipitate volume fraction*

The temporal evolution of the γ' precipitates was studied for the 2Nb2Re alloy after long term aging at 900 °C. Figs. 4(a-e) show the sequence of backscattered electron (BSE) images acquired from the grains orientated near to the [100] cubic direction for the 2Nb2Re samples aged for 50, 200, 300, 500, 1000 hours. All 2Nb2Re samples exhibit a γ/γ' microstructure without any evolution of other deleterious phases even at grain boundaries. Additionally, we observe no change in the γ' precipitate morphology even up to 1000 hours of aging at 900°C. However, with aging time, γ' precipitates become coarser and the inter-precipitate distance increases. Fig. 5(a) shows the variation in γ' precipitate volume fraction, an equivalent of area fraction, with aging time. For 50 hours aged sample, the γ' volume fraction ($\phi_t^{\gamma'}$) in 2Nb2Re is 53 ± 3%, which is similar to the value for 2Nb alloy [72] and also close to the equilibrium volume fraction ($\phi_{APT}^{\gamma'}$ = 54 ± 3%), as determined from the composition of the phases by using the lever rule based on APT composition measurements. This suggests that Re addition does not have an influence on the γ' volume fraction. However, the effect of Re is more pronounced during long term aging. The γ' volume fraction in the 2Nb2Re alloy does not change significantly with the aging time even after 1000 hours, the $\phi_{1000}^{\gamma'}$ measured to be ~ 48 ± 3%. In contrast, the γ' volume fraction in 2Nb alloy reduces to as low as ~ 25% after 1000 hours of aging at 900 °C due to continuous dissolution of γ' precipitates [72]. The similar dissolution is also seen in Ni-Al based superalloys where the long term aging at temperatures close to the solvus temperature leads to the dissolution of the γ' precipitates [75,76]. Thus, Re addition to 2Nb alloy plays a role of imparting microstructural stability by averting the γ' dissolution on long term aging up to 1000 hours at 900°C.

### 3.2.2 *γ' precipitate morphology*

From the BSE images in Fig. 4, we observe no change in the γ' precipitate morphology with the aging time till 1000 hours. They remain round-cornered cuboidal shaped precipitates. To quantify the effect of coalescence, coagulation or directional coarsening on the γ' precipitates, a parameter called "aspect ratio" has been defined. It is the ratio of the diameter of the largest circle that can envelop the precipitate touching at least two corners and the diameter of the smallest circle that can be inscribed inside the precipitate touching at least two points of the precipitate boundary. For a square or rectangular shape this corresponds to the ratio of the diagonal and the minimum separation of the two parallel faces. For a spherical precipitate, the 2D projection of the 3D objects makes the aspect ratio 1 while for the cubic precipitate it is $\sqrt{2}$ [28,77,78]. These are measured for precipitates (~ 500 numbers) from the [100] oriented grains of the 2Nb2Re samples aged at 900 °C. The probability distribution function (PDF) of aspect ratios with the aging time is plotted and is shown in the supplementary Figs. S2(a-d) for 2Nb2Re alloy. The PDF for all the aging times is asymmetric and skewed towards high values. Figure 5(b) shows the comparison of variation of mean aspect ratio (<ρ>) for the 2Nb2Re and the base 2Nb alloys. For the 2Nb2Re, the aspect ratio changes from a value of 1.3 for 50 hours aged sample to 1.2 after 200 hours and remains similar up to 1000 hours of aging at 900°C. The mode values of the PDF remain close to 1.1 for all aging times, which suggests that the coarsening of the γ' precipitates does not occur through the coagulation or coalescence mechanism. Additionally, the precipitates remained uniformly distributed without any alignment in any particular direction.

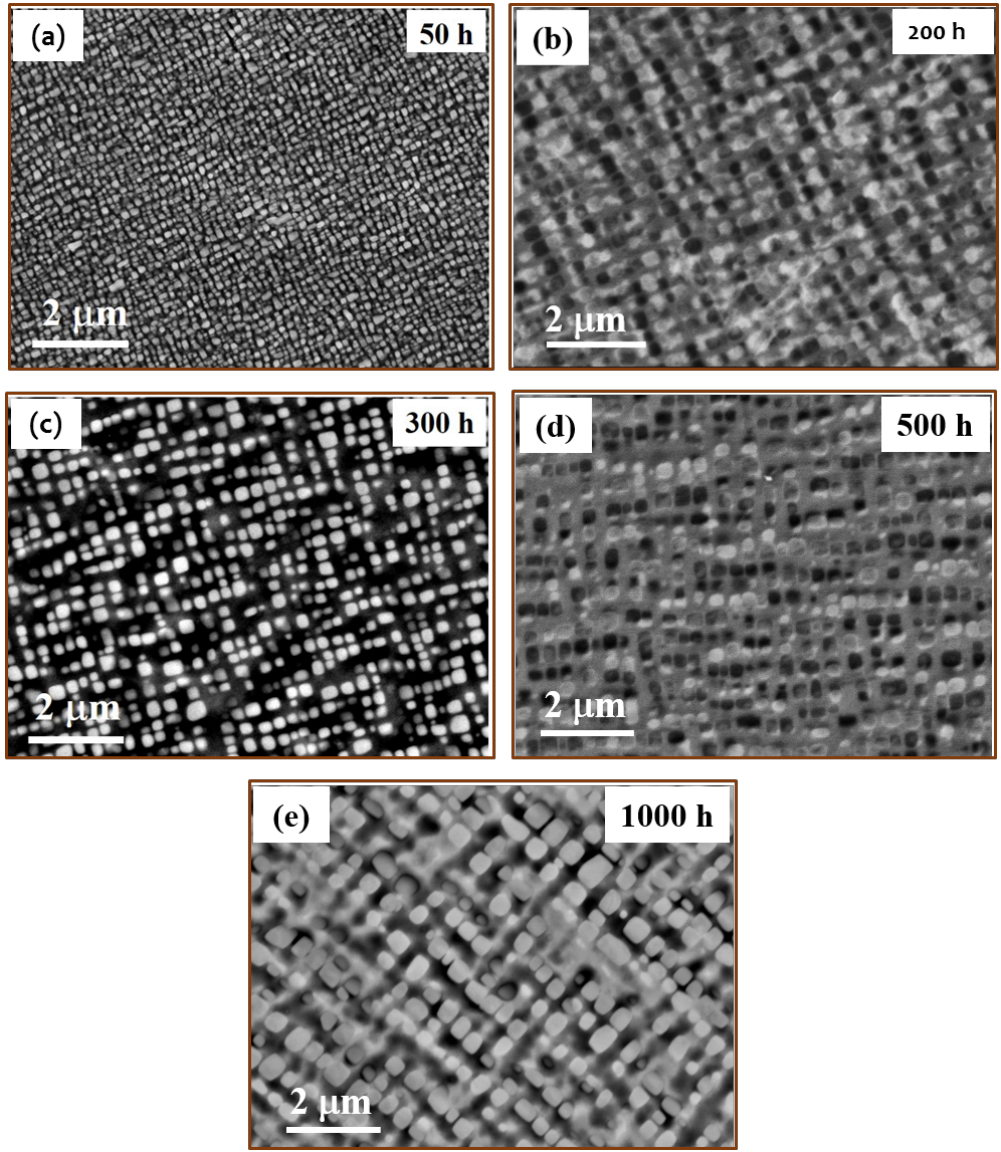

*Figure 4: (a-e) Backscattered SEM micrographs showing temporal evolution of γ/γ' microstructures in 2Nb2Re alloy subjected to aging at 900 °C for 50 to 1000 hours from the grains oriented near to [100] cubic direction.*

### 3.2.3 Average γ' precipitate size (<r>)

The change in the mean γ' precipitate size (<r>) with the aging time at 900°C for 2Nb2Re alloy is plotted in Fig. 5(c) and summarized in Table 3. <r> increases monotonously with the isothermal aging time. The histograms of precipitate size distributions (PSD) as a function of normalized precipitate radius (r/<r>) for different aging times are plotted and shown in the supplementary Figs. S3 (a-e). These histograms are consistent with a binomial (i.e. normal) distribution. During the early stages (50 and 200 hours), the PSD remains narrow with the maximum distribution achieved in the form of a sharp peak at the normalized r/<r> value close to 1. In contrast, as the γ' precipitates coarsen at later stages (300 to 1000 hours), the distribution becomes wider with an extended tail towards larger values of the normalized radius. The peak also transitions into a plateau within a specified range of r/<r> values. However, the center of the plateau still remains close to 1. The change in PSD of γ' precipitates from being narrow to broad during aging is also reflected in the standard deviation mentioned in Table 2. The mean edge-to-edge distance ($<\lambda_{e-e}(t)>$) between the γ' precipitates has also been calculated for each aging time using equation (5). The temporal change in the $<\lambda_{e-e}(t)>$ between the γ' precipitates, Figure 5 (d), shows continuous increase in the value with the aging time at 900 °C.

Table 2: *Temporal evolution of microstructural characteristics of γ' precipitates, namely precipitate mean size, maximum and minimum precipitate diameter (diameter of the largest and smallest circles) and aspect ratio*

| Time (h) | Mean size, (<r>) ± $\sigma_r$, nm | Standard Dev., $\sigma_{sd}$ | Maximum precipitate diameter, nm | Minimum precipitate diameter, nm | Aspect Ratio |
|---|---|---|---|---|---|
| 50 | 60.2 ± 0.6 | 14.0 | 121.9 ± 1.5 | 94.2 ± 1.0 | 1.3 |
| 200 | 93.1 ± 1.0 | 22.6 | 152.0 ± 2.1 | 180.4 ± 2.9 | 1.2 |
| 300 | 104.9 ± 1.1 | 29.5 | 205.5 ± 2.3 | 169.3 ± 1.9 | 1.2 |
| 500 | 122.8 ± 1.4 | 34.3 | 236.7 ± 3.0 | 201.3 ± 2.3 | 1.2 |
| 1000 | 160.1 ± 1.5 | 34.4 | 329.1 ± 3.6 | 269.6 ± 2.8 | 1.2 |

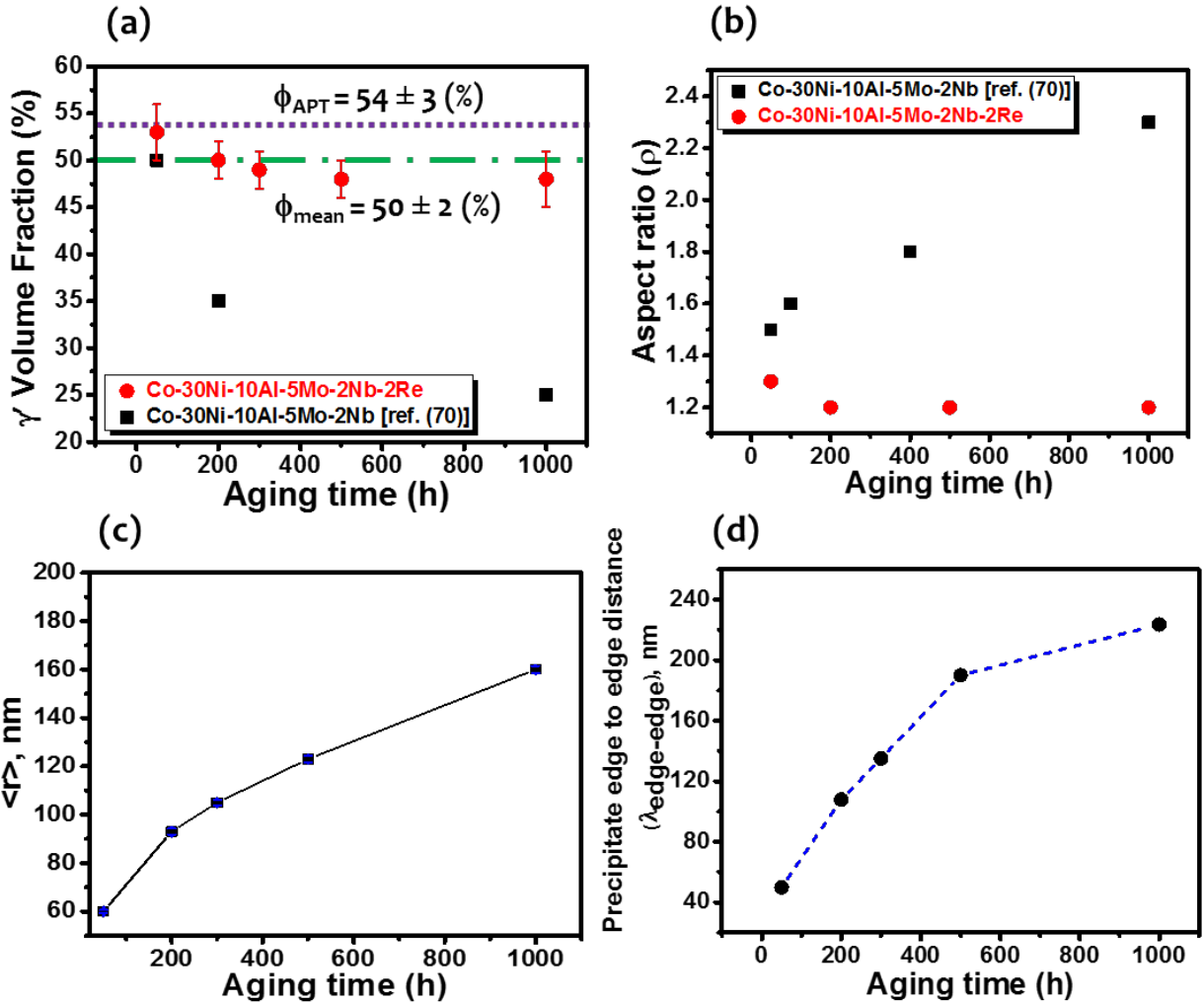

*Figure 5: (a) Change in the γ′ volume fraction of 2Nb2Re alloy during aging at 900 °C, as determined from the SEM images. The red dotted line at ϕ = 54 ± 3 % is the volume fraction as determined from APT composition using mass balance based lever rule. The green dashed at ϕ = 50 ± 3 % is the mean volume fraction. (b) Variation in the precipitate aspect ratio as a function of aging time at 900 °C for Co-30Ni-10Al-5Mo-2Nb-2Re (2Nb2Re) alloy and comparison with Co-30Ni-10Al-5Mo-2Nb (2Nb) alloy [ref. (70)]. (c) Plot showing the temporal evolution of precipitate size during aging at 900 °C (d) Change in mean edge to edge distance between precipitates [< $\lambda_{e-e}(t)$ >] for Co-30Ni-10Al-5Mo-2Nb-2Re (2Nb2Re) alloy during aging at 900 °C.*

### 3.2.4 *Compositional partitioning across the γ/γ′ interface*

The change in compositional partitioning of solutes across the γ/γ′ interface for 2Nb2Re alloy upon aging at 900°C from 50 to 1000 hours is obtained by APT. Supplementary Figs. S4 (a-c) show the reconstructions obtained with the γ/γ' interfaces delineated by iso-surface with a threshold composition 46 at.% Co for the 2Nb2Re alloy after 200, 500, and 1000 hours of aging. The composition profiles across the γ/γ' interface are also shown and compared. The variation in the compositions of γ, γ′ and the solute partitioning coefficients ($K_i$) are summarized in Table 3. The major effect of aging time can be observed in the partitioning coefficient value of Al ($k_{Al}$) that changes from 2.82 (50 hours) to 1.78 (200 hours) and remains similar up to 1000 hours. Additionally, the segregation of Mo and Re ahead of the γ/γ' interface reduces after 200 hours of aging indicating that the solute distribution in the γ matrix phase has become more uniform. Similar observations were made in Co-Al-W based alloys where the W solute segregated ahead of the interface in the γ matrix during early stage of precipitation and get reduced with the aging time [79].

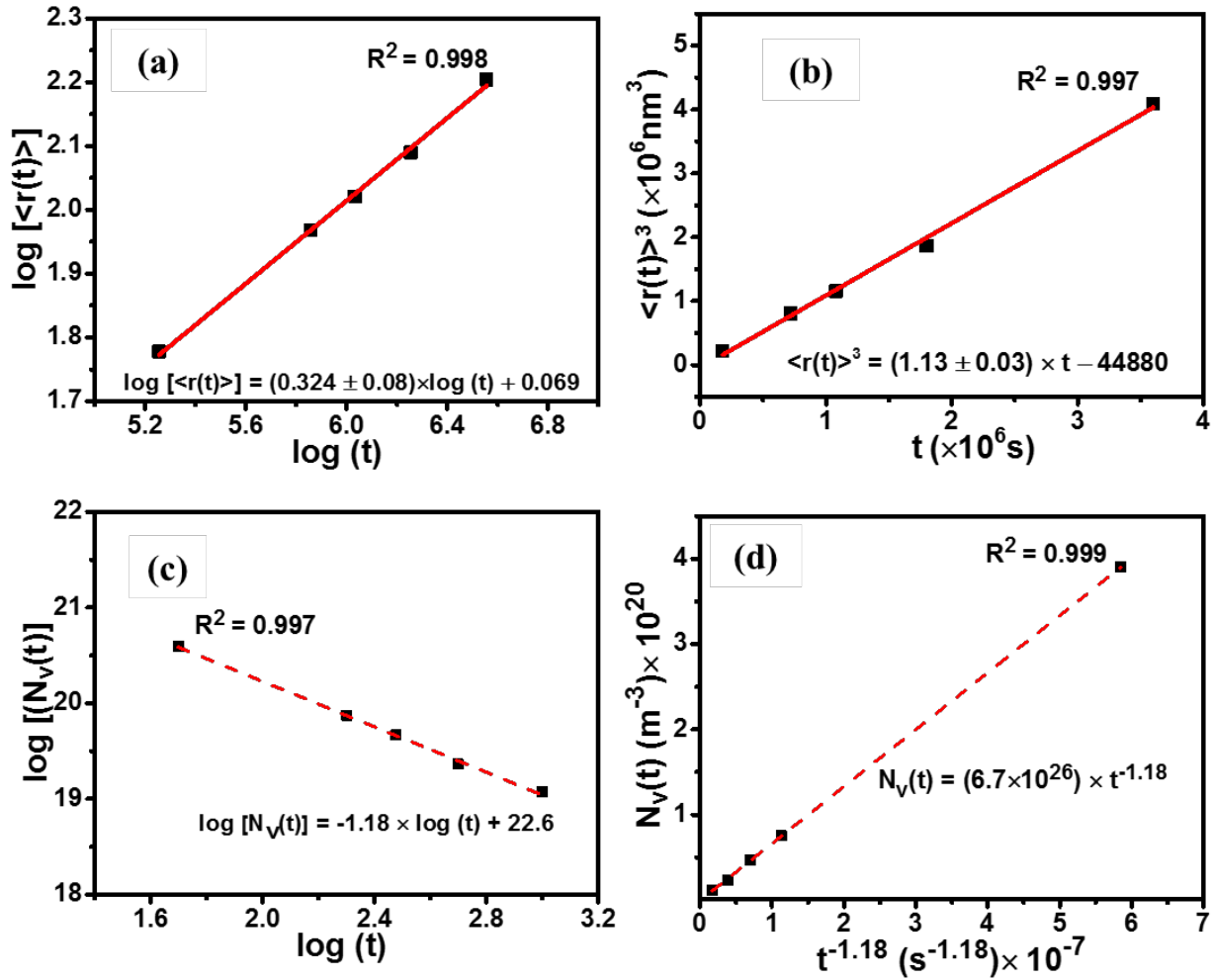

*Figure 6: (a) Temporal evolution of the γ' - precipitates during aging at 900 °C: (a) The plot between $\log \frac{[<r(t)>]}{[r<(0)>]}$ vs. log(t) shows temporal exponent of γ'-precipitate size evolution is equivalent to the exponent in classical LSW model and modified LSW models. (b) The plot between $[<r(t)>^3 - <r(0)>]^3$ vs. aging time (t) with fit to the experimental data using liner regression analysis gives coarsening rate constant ($K_r$). (a) The plot between $\log [N_v(t)]$ and log (t) gives temporal exponent, which is close to exponent in classical LSW model and modified LSW models. (b) The plot between precipitate number density ($N_v$) vs. aging time ($t^{-1.18}$) with linear fit using liner regression analysis gives coarsening rate constant ($K_v$).*

*Table 3: Compositions (at. %) of γ and γ' phase of Co-30Ni-10Al-5Mo-2Nb-2Re (2Nb2Re) alloy obtained from APT for 200h and 500h and 1000h aged alloy at 900 °C temperature.*

| Elements | 900 °C, 200 hours | | | 900 °C, 500 hours | | | 900 °C, 1000 hours | | |
|---|---|---|---|---|---|---|---|---|---|
| | $γ_p$ | $γ'_p$ | $K_i$ | $γ_p$ | $γ'_p$ | $K_i$ | $γ_p$ | $γ'_p$ | $K_i$ |
| Co | 55.40 ± 0.30 | 40.71 ± 0.30 | 0.73 | 56.02 ± 0.30 | 40.62 ± 0.50 | 0.73 | 55.07 ± 0.50 | 39.82 ± 0.50 | 0.72 |
| Ni | 28.24 ± 0.30 | 37.30 ± 0.20 | 1.32 | 27.50 ± 0.30 | 37.08 ± 0.50 | 1.35 | 27.85 ± 0.40 | 37.60 ± 0.40 | 1.35 |
| Al | 7.66 ± 0.10 | 13.63 ± 0.20 | 1.78 | 7.48 ± 0.20 | 13.67 ± 0.30 | 1.83 | 8.34 ± 0.30 | 14.57 ± 0.30 | 1.75 |
| Mo | 5.20 ± 0.20 | 4.36 ± 0.20 | 0.84 | 5.60 ± 0.20 | 4.77 ± 0.20 | 0.85 | 5.48 ± 0.20 | 4.88 ± 0.20 | 0.89 |
| Nb | 1.20 ± 0.10 | 3.06 ± 0.10 | 2.55 | 1.07 ± 0.10 | 2.88 ± 0.20 | 2.69 | 0.85 ± 0.10 | 2.14 ± 0.10 | 2.52 |
| Re | 2.30 ± 0.10 | 0.94 ± 0.10 | 0.41 | 2.33 ± 0.10 | 0.98 ± 0.10 | 0.42 | 2.41 ± 0.1 | 0.99 ± 0.10 | 0.41 |

3.2.5 *Estimation of coarsening exponent (n), rate constant ($K_r$) and Interfacial energy (σ)*

In the scenario of steady-state coarsening, the increase in mean precipitate size (<r(t)>) with the aging time (*t*) is given by equation 9. The value of the coarsening exponent (*n*) can be determined by the slope of the line from a plot between $\log \frac{[<r(t)>]}{[<r(0)>]}$ vs log (*t*). Fig. 6(a) shows the plot and the multiple linear regression analysis gives a coarsening exponent value (n) of 0.324 ± 0.08. This value is within the error equivalent to the value 0.33 predicted by the LSW model for binary alloys, the KV model for ternary alloys and the PV model for multicomponent alloy [63,70,71]. It indicates that the γ' coarsening rate is controlled by volume diffusion.

In Fig. 6(b), $[<r(t)>^3 - <r(0)>^3]$ is plotted as a function of the aging time (*t*). The coarsening rate constant ($K_r$), obtained by linear regression, is estimated to be ~ $1.13 \times 10^{-27}$ (m³/s).

The interfacial energy between the γ phase and the γ' precipitate in the 2Nb2Re alloy is calculated by using the experimental coarsening data and the modified LSW coarsening model given by Calderon et al. [69] for concentrated alloys. The γ/γ' interfacial energy (σ) can be estimated by assuming a value for some of the key parameters. First, the calculated molar volume of the precipitate ($V_m^{\gamma'}$) is 6.99 × 10$^{-6}$ m³/mole. Second, assuming Re is the slowest diffusing solute in the alloy and hence controls the coarsening rate, the diffusivity values of Re in fcc-Co in the temperature range of 1100 to 1300 °C were calculated by Neumeier et al. [56,80]. The extrapolation of this data to 900 °C yields the interdiffusion coefficient value (D) of 1.8 × 10$^{-18}$ m²/s. Using the equilibrium Re solute concentration values in the γ and γ', the interfacial energy (σ) can be estimated to be ~ 8.4 mJ/m². This value is similar to the estimated interfacial energy ~ 10 mJ/m² for Co-10Al-10W alloy at 900 °C, which is also calculated based on Calderon model.

### 3.2.6  γ' Precipitate number density (<$N_v$>)

The theoretical prediction for the γ' precipitate number density evolution with time is shown in Fig. 6(c). The slope of the plot between log [$N_v(t)$] and log ($t$) with linear regression analysis predicts the exponent value (*n*) to be ~ -1.18 ± 0.04. The temporal evolution of the γ' precipitate number density, which is calculated to decrease with $t^{-1.18 \pm 0.04}$, is in good agreement with theoretical value of $t^{-1}$ predicted by the LSW model and the PV model in the quasi stationary coarsening regime. Additionally, the slope of the plot between $N_v(t)$ and $t^{-1.18}$, as shown in Fig. 6(d), gives the amplitude ($\frac{0.21 \phi^{eq}}{K_r}$) value of 6.7 × 10$^{26}$ s/m³. The calculated amplitude value by keeping the estimated rate constant $K_r$ ~ 1.13 × 10$^{-27}$ (m³/s) is found to be ~ 1.0 × 10$^{26}$ s/m³, which is of the same order of magnitude as determined by the slope of the plot between $N_v(t)$ and $t^{-1.18}$. Thus, we show a reasonable agreement between rate constant values ($K_r$) obtained from the evolution of γ' precipitate size and the number density.

### 4. Discussion

In the γ/γ' Ni-Al based superalloys (2$^{nd}$ and 3$^{rd}$ generation), slow diffusing Re is a crucial alloying element to impart microstructural stability at high temperatures. The prime motivation behind this work is to evaluate the effect of Re in the new γ/γ' Co-Al based superalloys, in particular, the low density Co-Ni-Al-Mo-Nb based alloy (2Nb). The main results of the present study can be laid out through the following crucial points:

1) Re strongly partitions to the γ matrix with respect to the γ' precipitate and also shown to have influence in reversing the partitioning behavior of Mo solute in 2Nb2Re alloy.

2) Re addition led to a significant reduction of the γ/γ' lattice misfit by around ~ 40% that resulted in a change in the γ' morphology from cuboidal to round-cornered cuboidal precipitates.

3) The solvus of the 2Nb2Re alloy increased by ~ 10°C. Although, the density of the 2Nb2Re alloy increased to 8.78 gm/cc, the alloy exhibits similar 0.2% PS and superior specific 0.2% PS values even at high temperatures compared to Co-Al-W-based superalloy.

4) Re is shown to have significant influence on the γ' stability even up to 1000 hours of aging at 900°C in terms of γ' volume fraction, their morphology and their size distribution (PSD). We shall now discuss these results in the following sections.

4.1 *The γ' morphology, lattice misfit and solute partitioning across γ/γ' interface for 2Nb2Re alloy aged at 900° for 50 hours*

The equilibrium morphology of coherent γ′ precipitates in the γ matrix depends upon the two competing factors that contribute to the total free energy such that it achieves a minimum value. These factors are the γ/γ′ interfacial energy (chemical term, $\propto r^2$, r is the precipitate radius) and the elastic strain energy (strain term, $\propto r^3$) arising from the lattice misfit across the γ/γ′ interfaces. Here, the γ fcc matrix is elastically anisotropic and hence, the γ′ precipitates grow in the elastically soft [100] directions of the γ matrix, which results in a cuboidal shape of the γ' precipitates with a minimum elastic strain energy. At lower γ/γ' lattice misfit values, the effect of elastic strain energy on minimizing the total energy reduces and thus the γ' precipitates tend towards spherical shape that has a minimum interfacial energy. From the high-resolution X-ray diffraction at room temperature, the Re addition to the 2Nb alloy decreases the γ/γ' lattice misfit from 0.32% to 0.19% [72]. This leads to the change in the γ′ morphology from perfect cuboidal to round-cornered cuboidal shape. This morphological change in γ' shape is an outcome of the dominance of γ/γ' interfacial energy over the elastic strain energy in minimizing the total free energy. In Ni-Al based superalloys, the γ/γ′ lattice misfit is strongly sensitive to the partitioning of solutes across the γ/γ' interface [24,81]. The composition profile across the γ/γ' interface in Fig. 2(b) shows that the Re addition leads to the partitioning reversal of Mo from the γ' precipitate to the γ phase concomitantly with strong partitioning of Re to the γ phase. In the 2Nb alloy, Mo strongly partitions to γ' along with Nb and Al. Hence, the accommodation of large Mo and Re atoms in the γ phase of 2Nb2Re alloy result in an increase in the γ lattice parameter. This is consistent with the smaller γ lattice parameter (3.575 Å) for the 2Nb alloy compared to the present 2Nb2Re alloy (3.587 Å). This increase in the γ lattice parameter decreases the overall γ/γ' lattice misfit according to the equation (8). Even in the model Ni-Cr-Al superalloy, Re is known to stabilize the spheroidal morphology of the γ' precipitates. In contrast, Re addition has no apparent effect on the morphology of γ' precipitate in the Co-Al-W based superalloys [54].

In addition to the Re partitioning to the γ phase, we also have an evidence of segregation of Re and Mo to the γ/γ' interface in the 50 hours aged sample, Fig. 2(c). Similar segregation effects of Re at the γ/γ' interfaces were observed in the multicomponent Ni-Al based (René N6) [82,83] and of Mo in Co-Al-W based superalloys [49]. These solute segregations contribute to reducing the interfacial energy of the γ/γ' interfaces and also are indicative of the matrix diffusion controlled growth of the γ' precipitates in the γ matrix. The segregation of solutes at the γ/γ' interfaces can be quantified using Gibbsian interfacial excess (IE) of the respective solute ($\Gamma_i$). To accurately calculate the IE from APT datasets, we developed a novel data treatment method, which is described in a separate publication [61]. Fig. 7 shows the final 2D quantitative IE maps of (a) Re and (b) Mo along the γ/γ' interface marked by the semi-transparent cube in Fig. 2(a). Both Re and Mo exhibit a certain degree of segregation to the γ/γ' interface. However, the distributions of their IE along the interface are inhomogeneous. The $\Gamma_{Re}$ value ranges from minimum value of 0.8 to maximum value 9.6 atoms.nm$^{-2}$. The $\Gamma_{Mo}$ value ranges from minimum value of 0 to maximum value 7.6 atoms.nm$^{-2}$. These variations can probably be attributed to the effect of local curvature of the γ/γ' interface. To closely compare the difference between the segregation behavior of Re and Mo at the γ/γ' interface, histograms of their IE values are plotted in Fig. 7(c). In contrast to Mo, the segregation of Re is more pronounced. Most of the regions at the γ/γ' interface showed excess $\Gamma_{Re}$ value, which indicates that in the 2Nb2Re alloy, Re contributes more in reducing the γ/γ' interfacial energy.

*4.2 γ' Coarsening mechanism on long term aging at 900°C (up to 1000 hours)*

We have shown in section 3.2, the detailed evolution of 1) γ' precipitate morphology by calculating the probability distribution function (PDF) of aspect ratios, 2) γ' precipitate size (<r>) by calculating the precipitate size distribution (PSD) as a function of normalize precipitate radius (r/<r>), and 3) the change in mean edge to edge distance ($<\lambda_{e-e}(t)>$) between the γ' precipitates with the aging time up to 1000 hours at 900°C. All these measurements show that the coarsening of γ' precipitates in the 2Nb2Re alloy occurs by evaporation-condensation (EC) mechanism which is driven by the minimization of total interfacial energy through reduction of total interfacial area. This lead to the growth of bigger particles at the expense of shrinking particles [64].

Further, the mode values of PDF remain near 1.1 when the sample is aged for 1000 hours at 900°C. The γ' precipitates in this case also coarsen without any directional alignment in the γ matrix phase. This is in contrast to the case of 2Nb base alloy, where, extensive alignment of γ' precipitates along <100> directions and their growth through coagulation/coalescence could be observed [72]. The cuboidal-shaped γ' precipitates (with aspect ratio ~ 1.5 ), in the 2Nb alloy, transforms to a rod morphology (aspect ratio ~ 2.3, as calculated from the Ref. [72])) after 1000 hours of aging at 900°C. Additionally, in 2Nb2Re alloy, the temporal evolution of the mean precipitate size <r(t)> (see Table 3 and Fig. 5(c)) shows a continuous increase with aging time that occurs concomitantly with a decrease in the average precipitate number density, $N_v(t)$. The PSD in supplementary Figs. S3(a-e) show that the peak of the normal distribution is achieved at the normalized radius value of ($\frac{r}{<r>}$) = 1.0 for all the aging times. This value indicates that a steady-state coarsening regime has been achieved after 50 hours of aging at 900°C. The PSD after 50 hours of aging is narrower and becomes wider with time up to 1000 hours that further supports that the precipitate coarsening is not occurring through the coalesce and coagulation mechanism. This is also supported by the continuous increase in the γ' precipitate edge to edge $<\lambda_{e-e}(t)>$ distance with the aging time at 900°C. The calculated exponent value during the temporal evolution of precipitate number density is -1.18 ± 0.04, which is in good agreement with the theoretical value of -1 as predicted by various modified LSW models for non-dilute multicomponent systems. The slight variation suggests that the quasi-stationary state of coarsening regime has been reached.

The coarsening of γ' by the coagulation and coalescence mechanism has been established in Co-Al-W based and Ni-Al based superalloys [84] [85–89]. The characteristic alignment of the γ' precipitates along a particular direction during coarsening is referred to as γ' rafting [90–92]. Rafting can occur with or without external stress at high temperatures. The γ' rafting along elastically soft <100> type directions during annealing (without external stress) has been shown in Ru containing Co-Al-W superalloys [84] and even in W containing Ni-Cr-Al based superalloys [85–89]. During the γ' coarsening process, in Ni-Al-Cr system, the γ' precipitate morphology changes from spheroidal to cuboidal shape and subsequently these precipitates align themselves along the <100> type directions in the γ matrix phase [93,94]. Hence, the addition of Re to the base 2Nb alloy prohibits the coarsening of the γ' precipitates by coagulation/coalescence mechanism and thus avoids the formation of elongated rod shaped γ' precipitates during prolonged aging up to 1000 hours at 900°C. This is attributed to the reduced lattice misfit across the γ/γ' that led to minimization of the elastic anisotropy.

4.3 *Coarsening exponent, rate constant and γ/γ' interfacial energy*

The coarsening exponent (n) can be estimated from the slope of the plot between $log \frac{[<r(t)>]}{[<r(0)>]}$ vs. log(t), as shown in Fig. 6(a). The value is found to be 0.324 ± 0.08, which is close to the value of 0.33 predicted for LSW, KV and PV models. This indicates the coarsening of the γ' precipitates in the γ matrix is diffusion controlled. The experimentally calculated coarsening rate constant ($K_r$) for the γ' precipitates in 2Nb2Re alloy at 900 °C is 1.13 × 10$^{-27}$ (m$^3$/s). This is comparable to the coarsening rate constant of 1.41 × 10$^{-27}$ (m$^3$/s) reported for γ' precipitates in W containing model Co-10Al-10W system [79], 0.55 × 10$^{-27}$ (m$^3$/s) for Co-8.8Al-7.3W system [84], 0.27 × 10$^{-27}$ (m$^3$/s) for Ru containing Co-9.7Al-7.1W-2.1Ru system [84] and 0.67 × 10$^{-27}$ (m$^3$/s) for Co-9.1Al-7 W system [95]. Additionally, the obtained value of $K_r$ for γ' in the 2Nb2Re alloy is lower than the $K_r$ values for two recently reported low density W containing Co-Al based superalloys Co-10Ni-7Al-4Ti-2W-3Mo-1Nb-1Ta and Co-30Ni-7Al-4Ti-2W-3Mo-1Nb-1Ta, which have the values of 3.9× 10$^{-27}$ (m$^3$/s) and 3.8× 10$^{-27}$ (m$^3$/s), respectively. The calculated interfacial energy (σ) across γ/γ' interfaces for the 2Nb2Re alloy at 900 °C is 8.4 mJ/m$^2$. This is lower than the reported values of 10 mJ/m$^2$ to 48 mJ/m$^2$ for Co-Al-W-based high density superalloys containing 7 to 10 at.% W [49,79,95]. The different values of $K_r$ and σ in our experiments reflects the influence of Re addition to the base 2Nb alloy on the long-term microstructural stability. The slower coarsening rate is attributed to the slow diffusivity of Re in the γ matrix and γ/γ' misfit reduction resulting in a weakening of the effect of elastic strain energy across the γ/γ' interfaces.

As mentioned earlier, in section 4.1, the enrichment of Re and Mo near the γ/γ' interface could be observed for the 2Nb2Re alloy after aging at 900°C for 50 hours. During the growth of the γ' precipitates, the Re and Mo atoms get ejected from the γ' precipitates and diffuse towards the γ matrix. This interfacial segregation at the γ/γ' interface results from the sluggish diffusivities of heavy Re and Mo atoms in the γ matrix as the system aims to reach a local equilibrium. The sluggish kinetics prevents the system from reaching a global equilibrium in the initial stages of aging. Only after long term aging at 900 °C is the equilibrium Re and Mo compositions achieved at the matrix–precipitate interface. This is evidenced in the composition profiles of Re and Mo across the γ/γ' interface that show no segregation effects for samples aged > 200 hours at 900°C (see supplementary Fig. S5). However, the γ' precipitate aspect ratio ($\rho$) remains ~ 1.2 even after the aging for 1000 hours at 900°C and hence, the γ' precipitate morphology remains unaltered for 2Nb2Re alloy. Thus, the rearrangement of Re and Mo solutes across the γ/γ' interface during long term aging, have negligible influence on the γ' morphology.

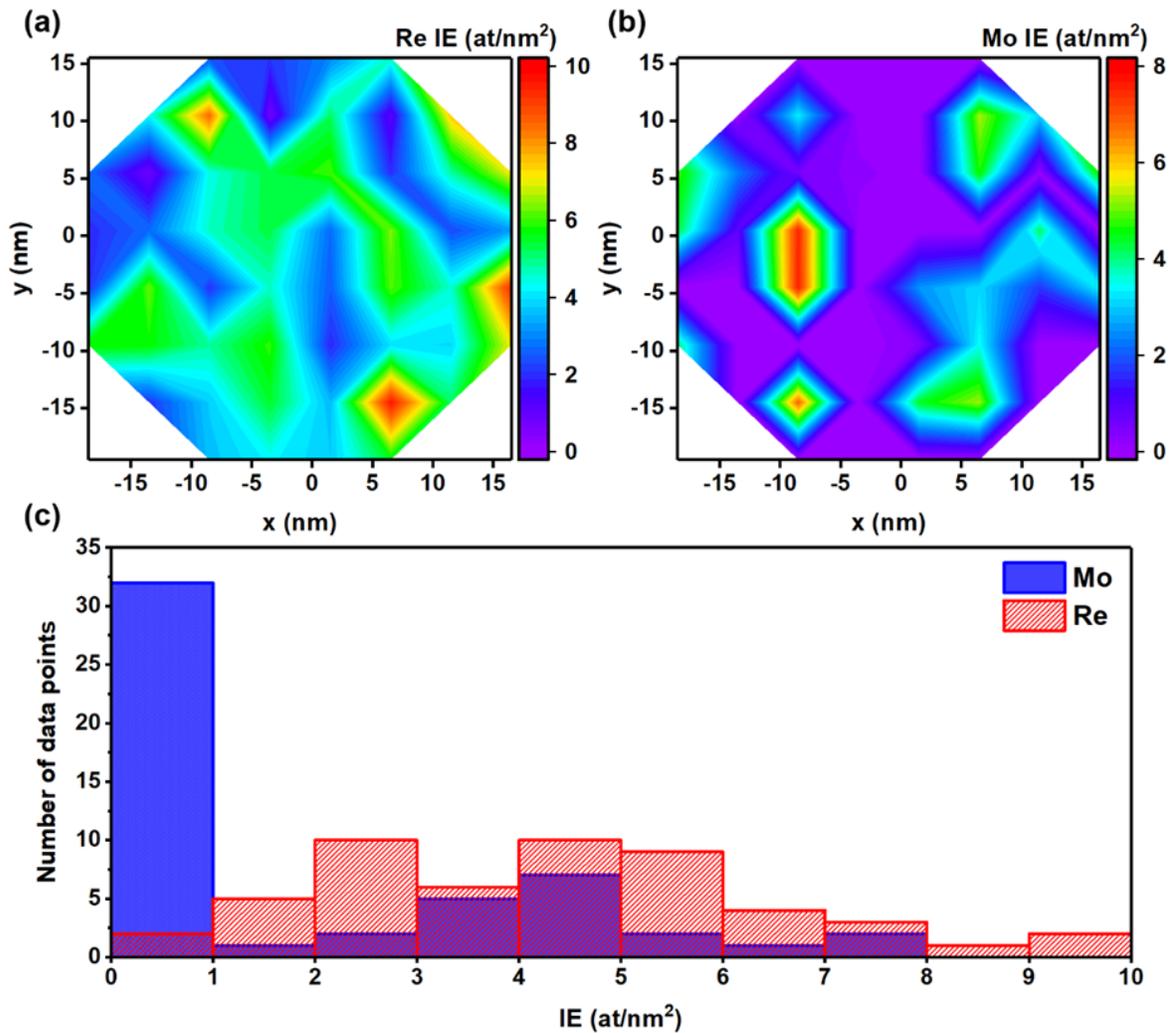

*Figure 7: 2D quantitative maps showing the variations of Gibbsian interfacial excess (IE) of the solutes (a) Re and (b) Mo along the γ/γ′ interface marked by the semi-transparent cube in Fig. 2(a). To compare the difference between their distributions at the interface, histograms of their IE values are shown in (c).*

4.4 *Mechanical properties*

The strength of the 2Nb2Re alloy remains similar to the base 2Nb and Co-Al-W based superalloy up to a temperature of 570°C. Beyond this, the 0.2% proof stress of Re containing alloy have comparatively higher values. This suggest that the Re addition influences the high temperature mechanical behavior of these Co-Al based superalloys. In Ni-Al based CMSX-2 superalloy, addition of 6wt.% Re (CMSX-10) increases the creep rupture life from 250 hours to 2500 hours at a temperature of 750°C and a stress level of 500 MPa [2]. This significant enhancement in high temperature properties is termed as "Re effect" [15]. Earlier research [96,97] proposed that the possible formation of Re nano-clusters in the γ matrix phase could hinder dislocation motion. However, evidence of Re clusters were not found by APT [98] and extended X-ray absorption fine-structure spectroscopy [99]. Recent investigations [100,101] show that the Re atoms segregates to the interfacial dislocation cores as Cottrell atmosphere and is responsible for stopping dislocation motion in particular at high temperatures. Further experiments by APT on creep deformed CMSX-4 alloy (3wt.% Re) show a significant segregation of Re atoms (~ 2at.%) at the core of dislocations present inside the γ' phase [102]. This indicates that the movement of dislocations during high temperature deformation experiences a Cottrell drag force by the slow diffusing Re atoms. In the 2Nb2Re alloy, we expect similar effect of Re on the motion of dislocations during high temperature deformation. This will be our future goal for detailed analysis on "Re effect" in high temperature deformation behavior of Co-Al based superalloys.

## 5. Conclusions

In summary, this work elucidates the pronounced effect of Re on microstructural evolution of new Co-30Ni-10Al-5Mo-2Nb based alloy. Based on the obtained results the following major conclusion can be drawn:

1. The Re addition to the base Co-30Ni-10Al-5Mo-2Nb (2Nb) changes the γ' precipitates morphology from cuboidal to round-cornered cuboidal shape by reducing the γ/γ' lattice misfit i.e. from high misfit (+0.32%) to low misfit (+0.19%) [72]. The γ' solvus temperature increases by 10 °C and have comparable high temperature mechanical properties to other high-density Co-Al-W based superalloys.
2. In 2Nb2Re alloy, the Re strongly partitions to the γ matrix ($K_{Re} = 0.34$) and reverses the partitioning of Mo from γ' precipitates to the γ matrix. In 900°C, 50 hours aged sample, Re is shown to have inhomogeneous segregation at the γ/γ' interface with a Gibbsian excess value, $\Gamma_{Re}$ ranging from 0.8 to 9.6 atom.nm$^{-2}$. We also observe interfacial excess of Mo solute in a few regions of the γ/γ' interface. These segregation effects reduce after 200 hours of aging at 900°C.
3. During prolonged aging at 900°C, the coarsening of γ' precipitates in 2Nb2Re alloy occurs by evaporation-condensation (EC) mechanism. In contrast, the γ' precipitates in the base 2Nb alloy show coarsening through coagulation/coalescence mechanism with extensive directional alignment of γ' along <100> directions [72].
4. The coarsening rate constant value ($K_r$) for the γ' precipitates, as predicted by the modified LSW model of non-dilute solution, estimated to be $1.13 \times 10^{-27}$ (m$^3$/s), which is comparable to the high-density Co-Al-W based superalloys. The γ/γ' interfacial energy for 2Nb2Re alloy is estimated to be ~ 8.4 mJ/m$^2$ at 900 °C. This value is lower than the values for other Ni-Al based and Co-Al-W based superalloys (10 mJ/m$^2$ to 48 mJ/m$^2$).

**Acknowledgements**

The authors would like to acknowledge the microscopy facility available at the Advanced Facility for Microscopy and Microanalysis (AFMM) center, Indian Institute of Science, Bangalore. KC is grateful for the financial support from **Department of Science and Technology (DST)** in the form of **J.C. Bose national fellowship**. KC also acknowledges the Gas Turbine Materials and Processes (GTMAP) programme of **Aeronautics Research and Development Board, DRDO** for the financial support. The authors are grateful to U. Tezins and A. Sturm for their technical support of the atom probe tomography and focused ion beam facilities at the Max-Planck-Institut für Eisenforschung. The authors also acknowledge Prof. Rajarshi Banerjee for valuable discussion. SKM acknowledges financial support from the **Alexander von Humboldt Foundation**.